# Implementation of radiating aperture field distribution using tensorial metasurfaces

Mounir Teniou, Hélène Roussel, Nicolas Capet, Gérard-Pascal Piau and Massimiliano Casaletti

*Abstract*—**This paper deals with the design of tensorial modulated metasurfaces able to implement a general radiating aperture field distribution. A new aperture synthesis approach is introduced, based on local holography and variable impedance modulation. In particular, it is shown that tensorial metasurfaces can be used to generate general radiating distribution (phase and amplitude). In addition, a step by step algorithm is presented. In order to validate the method, several solutions are presented at 20 GHz which implement aperture distributions able to radiate different beams with general polarization.**

## I. INTRODUCTION

In recent years, metamaterials have become an appealing subject of research. They are synthetic materials that have exotic properties that cannot be found in nature: double negative materials, negative index materials, left-handed materials… Metasurfaces are the equivalent of metamaterials in the case of 2D structures. The properties of these surfaces are described in terms of tensorial or scalar surface impedances (analogous to the constitutive parameters for volumetric metamaterials). Metasurfaces have been recently used in many applications like holographic antennas [1], leaky-wave antennas [2]-[4],[19], planar lenses [5]-[7], polarization convertors [8],[9], orbital angular momentum communication [10] or transformation optics [11]-[12].

All these works are based on the propagation properties of waves over a sinusoidally modulated surface impedance [13]. By choosing an appropriate modulated surface impedance, it is possible to control the propagation of SW along a surface or to obtain the transition from SW to leaky wave (LW) modes in order to realize antennas [16],[17].

Surfaces composed of sub-wavelength printed elements over grounded dielectric slabs were largely used in order to obtain modulated scalar impedances by locally changing the dimensions of the elements [1]-[15]. Symmetric elements are used to produce scalar impedances [1]-[3],[18],[19] while asymmetric elements can implement tensorial impedances by creating cross-pol field components [1],[20]-[22].

Scalar metasurface antennas can produce general polarized beams [1]-[3],[18],[19]. However, the direction of the radiating aperture field (or the equivalent surface current) is dictated by the source [19]. This latter aspect limits the number of possible aperture field distributions that can be implemented.

Recently, tensorial metasurfaces were successfully used in antenna design that can radiate CP waves [1],[20] and isoflux shaped beam antennas for space applications [20],[21].

The additional degrees of freedom offered by tensorial metasurfaces could be used to overcome the limits of scalar solution by generalizing the procedure presented in [19].

Our objective is to propose a systematic procedure for the design of metasurface antennas capable of implementing a general aperture field distribution (amplitude, phase and direction). The principal novelty of this approach is the independent control of the generated aperture field components. This important aspect (critical for general aperture implementation), is achieved by introducing independent modulations of the impedance tensorial components and a new exact holographic formulation. Moreover, average impedance variation along the propagation direction is introduced in order to compensate the spreading factor of the incident field and prevent undesired reflections.

This paper is structured as follows. Section II summarizes the basic properties of wave propagation over tensorial metasurfaces. Section III introduces the new aperture synthesis approach, based on the holography principle and variable impedance modulation. Section IV provides the theoretical results for an incident cylindrical wave (coaxial excitation). Section V presents a step by step algorithm for the design of tensorial metasurface antennas. Finally, some examples of aperture distributions implementation are presented in Section VI. Conclusions are drawn in Section VII.

## II. BACKGROUND

The surface impedance $\underline{\underline{\mathbf{Z}}}_s$ is defined as the ratio between tangential electric ($\mathbf{E}_t$) and the magnetic ($\mathbf{H}_t$) fields at the surface boundary $S$ for a particular wavevector ($\mathbf{k}^{sw}$):

$$\mathbf{E}_t\left(\boldsymbol{\rho}'\right)\Big|_{\boldsymbol{\rho}'\in S} = \underline{\underline{\mathbf{Z}}}_s\left(\mathbf{k}^{sw}\right)\hat{\mathbf{n}}\times\mathbf{H}_t\left(\boldsymbol{\rho}'\right)\Big|_{\boldsymbol{\rho}'\in S} = \underline{\underline{\mathbf{Z}}}_s\mathbf{J}\left(\boldsymbol{\rho}'\right) \qquad (1)$$

where $\boldsymbol{\rho'} = x'\hat{\mathbf{x}} + y'\hat{\mathbf{y}}$ is a point on the antenna surface, $\hat{\mathbf{n}}$ is the vector normal to the surface, $S$ and $\mathbf{J}(\mathbf{r}) = \hat{\mathbf{n}} \times \mathbf{H}_t(\mathbf{r})\big|_{\mathbf{r} \in S}$ is the equivalent surface current.

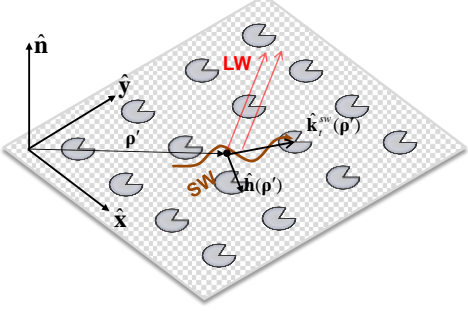

Fig. 1. Metasurface geometry and leaky wave generation

Since we deal with surface waves (discrete spectrum), the impedance (1) is defined only for a discrete set of wavenumbers, namely:

$$\mathbf{k}^{sw} = k_t^{sw}(n, \omega, \varphi)\hat{\mathbf{k}}_t^{sw} = k_t^{sw}(n, \omega, \varphi)\big(\hat{\mathbf{x}}\cos(\varphi) + \hat{\mathbf{y}}\sin(\varphi)\big) \quad (2)$$

where $k_{sw}$ is the wavenumber associated to the $n$-th mode propagating along the direction $\hat{\mathbf{k}}_t^{sw}$ defined by the angle $\varphi$ at the angular velocity $\omega$.

The impedance boundary condition can be implemented using different approaches, however, in any case it takes into account all of the interaction effects between the wave and the structure/media below the plane defined by z=0.

We will focus our attention on the dominant mode (n=0) propagating on reciprocal lossless metasurfaces. Conservation of energy (anti-Hermitian impedance tensor $\underline{\mathbf{Z_s}}^{\dagger} = -\underline{\mathbf{Z_s}}$ ) in conjuction with reciprocity imply that $\underline{\mathbf{Z_s}}$ is a purely imaginary symmetric tensor [23].

Thus, in a general 2D orthogonal reference system placed on S, equation (1) is written as

$$\begin{bmatrix} E_1 \\ E_2 \end{bmatrix} = j \begin{bmatrix} X_{11} & X_{12} \\ X_{12} & X_{22} \end{bmatrix} \begin{bmatrix} -H_2 \\ H_1 \end{bmatrix} = j \begin{bmatrix} X_{11} & X_{12} \\ X_{12} & X_{22} \end{bmatrix} \begin{bmatrix} J_1 \\ J_2 \end{bmatrix}. \quad (3)$$

For antenna applications, metasurfaces are used to convert an incident surface wave (SW), launched by a feeder, into a leaky wave (LW) mode, using a locally-periodic modulation [1]-[3],[14]-[21].

In the following, we suppose that the incident SW is propagating along a general direction $\hat{\mathbf{k}}_t^{sw}$ on a metasurface placed in the x-y plane as shown in Fig. 1.

For tensorial surface impedances, the dominant SW mode is a hybrid EH mode (more details can be found in the appendix). The magnetic field on the surface is assumed to be of the general form

$$\mathbf{H}_t(\boldsymbol{\rho'}) = A(\boldsymbol{\rho'})e^{-j\mathbf{k}^{sw}(\boldsymbol{\rho'}) \cdot \boldsymbol{\rho'}} \hat{\mathbf{h}}(\boldsymbol{\rho'}) \quad (4)$$

where A is the field amplitude, $\hat{\mathbf{h}}$ is the magnetic field polarization unit vector, $\mathbf{k}^{sw} = k_t^{sw} \hat{\mathbf{k}}_t^{sw}$ is the wave vector and the propagation constant $k_t^{sw}$ is found by solving the dispersion equation (see appendix (A5) and (A6)).

Let us consider a small modulation of the surface reactance components along the propagation direction of the form

$$X_{ij}(x) = \bar{X}_{ij}\left[1 + M_{ij}\cos\left(\frac{2\pi}{p_{ij}}x\right)\right] \quad (5)$$

where $\bar{X}_{ij}$, $M_{ij}$ and $p_{ij}$ are the average reactance, the modulation index and the periodicity of the $ij$ component, respectively. It should be noted that the periodicities of the reactance components are not necessarily identical $p_{ij} \neq p_{kl}$.

Under the assumption of small perturbation of the average impedance values ( $M_{ij} < 1$), eq.(4) can still be considered as a valid approximation of the magnetic field above the modulated impedance [1] [17], [22]. Thus, a first order estimation of the electric field with respect to small parameters $M_{ij}$ can be obtained by replacing (4), (5) into (3) and writing the cosine term of (5) in the exponential form, leading to :

$$\begin{cases} E_1 = j(\bar{X}_{11}J_1 + \bar{X}_{12}J_2 + \dfrac{M_{11}}{2}\bar{X}_{11}J_1 e^{-j\frac{2\pi}{p_{11}}x} + \dfrac{M_{12}}{2}\bar{X}_{12}J_2 e^{-j\frac{2\pi}{p_{12}}x} \\ \qquad + \dfrac{M_{11}}{2}\bar{X}_{11}J_1 e^{j\frac{2\pi}{p_{11}}x} + \dfrac{M_{12}}{2}\bar{X}_{12}J_2 e^{j\frac{2\pi}{p_{12}}x}) \\[2mm] E_2 = j(\bar{X}_{12}J_1 + \bar{X}_{22}J_2 + \dfrac{M_{12}}{2}\bar{X}_{12}J_1 e^{-j\frac{2\pi}{p_{12}}x} + \dfrac{M_{22}}{2}\bar{X}_{22}J_2 e^{-j\frac{2\pi}{p_{22}}x} \\ \qquad + \dfrac{M_{21}}{2}\bar{X}_{21}J_1 e^{j\frac{2\pi}{p_{21}}x} + \dfrac{M_{22}}{2}\bar{X}_{22}J_2 e^{j\frac{2\pi}{p_{22}}x}) \end{cases} \quad (6)$$

For each component of the electric field, the first line represents non radiative surface waves, while the terms of the second line can be leaky waves if $k_t^{sw} - 2\pi/p_{ij} < k_0$.

As can be seen in eq.(6), the phase of the excited leaky waves can be controlled by the periodicity of the reactance $p_{ij}$ while the amplitude is proportional to the product $\bar{X}_{ij}M_{ij}$.

## III. FORMULATION

This section presents a procedure to control the generated LW component of the field (6), namely

$$\mathbf{E}_t^{LW} = \begin{bmatrix} E_1^{LW} \\ E_2^{LW} \end{bmatrix} = j \begin{bmatrix} \dfrac{M_{11}}{2}\bar{X}_{11}J_1 e^{-j\frac{2\pi}{p_{11}}x} + \dfrac{M_{12}}{2}\bar{X}_{12}J_2 e^{-j\frac{2\pi}{p_{12}}x} \\[2mm] \dfrac{M_{21}}{2}\bar{X}_{21}J_1 e^{-j\frac{2\pi}{p_{21}}x} + \dfrac{M_{22}}{2}\bar{X}_{22}J_2 e^{-j\frac{2\pi}{p_{22}}x} \end{bmatrix} \quad (7)$$

in order to implement a desired objective radiating aperture field distribution:

$$\mathbf{E}_t^{obj} = \begin{bmatrix} E_1^{obj} \\ E_2^{obj} \end{bmatrix} = \begin{bmatrix} \left|E_1^{obj}\right|e^{j\arg(E_1^{obj})} \\ \left|E_2^{obj}\right|e^{j\arg(E_2^{obj})} \end{bmatrix} \quad (8)$$

### A. Phase control

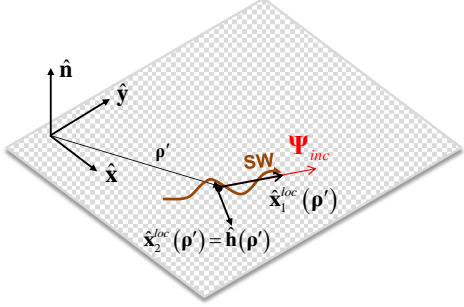

Fig. 2. Local framework definition.

Phase synthesis is based on a new holographic principle that is adapted to surface to leaky wave transitions. This technique enables the exact phase match of the objective field.

Two approaches may be adopted in order to obtain the desired phase distribution on the surface:

a) For each component, equating the phase of each individual term in (7) with the phase of (8).

b) For each component, equating the phase of the sum of the terms in (7) with the phase of (8).

In the first approach, imposing the same phase between each term of (7) with (8) leads to the following condition for the impedance periodicities:

$$p_{11} = \frac{2\pi}{\arg(J_1) + \Pi_{11} - \arg(E_1^{obj})}, \ p_{12} = \frac{2\pi}{\arg(J_2) + \Pi_{12} - \arg(E_1^{obj})}$$

$$p_{21} = \frac{2\pi}{\arg(J_1) + \Pi_{21} - \arg(E_2^{obj})}, \ p_{22} = \frac{2\pi}{\arg(J_2) + \Pi_{22} - \arg(E_2^{obj})}$$

(9)

where $\Pi_{ij} = \pi\left(\text{sgn}\left(\bar{X}_{ij}\right) - 1\right)/2$ and $\Pi_{12} = \Pi_{21}$.

The tensor impedance (5) with the periodicities given by (9) can be written in compact form if all of the average impedances $\bar{X}_{ij}$ have the same sign. In fact, by defining the *incident wave* $\boldsymbol{\Psi}_{inc}$ as the phase of the current $\mathbf{J}$, and the *objective wave* $\boldsymbol{\Psi}_{obj}$ as the phase terms of the objective electric field (8), namely:

$$\boldsymbol{\Psi}_{inc} = \begin{bmatrix} \Psi_1^{inc} \\ \Psi_2^{inc} \end{bmatrix} = \begin{bmatrix} e^{j\arg(J_1)} \\ e^{j\arg(J_2)} \end{bmatrix}, \ \boldsymbol{\Psi}_{obj} = \begin{bmatrix} \Psi_1^{obj} \\ \Psi_2^{obj} \end{bmatrix} = \begin{bmatrix} e^{j\arg(E_1^{obj})} \\ e^{j\arg(E_2^{obj})} \end{bmatrix}$$

(10)

the modulated tensorial impedance is obtained as:

$$\underline{\underline{\mathbf{Z}}}_s = j\underline{\underline{\mathbf{X}}} + j\underline{\underline{\mathbf{X}}} \circ \underline{\underline{\mathbf{M}}} \, \Im\left(\boldsymbol{\Psi}_{obj} \otimes \boldsymbol{\Psi}_{inc}^*\right)$$

(11)

where $\underline{\underline{\mathbf{X}}}$ is the average reactance tensor. $\underline{\underline{\mathbf{M}}}$ is the modulation index matrice, $\otimes$ and $\circ$ are the outer and the Hadamard products respectively.

As stated in Section II, $\underline{\underline{\mathbf{Z}}}_s$ must be anti-Hermitian. This condition is not guaranteed by (11) as the off diagonal impedance elements depend on field parameters that can not be controled a priori (incident field and objective field). This problem arises because we are trying to impose the 4 conditions in (9) with only 3 degrees of freedom given by the impedance tensor.

A possible solution is to decompose the tensor (11) as

$$\underline{\underline{\mathbf{Z}}}_s = \underline{\underline{\mathbf{H}}} + \underline{\underline{\mathbf{A}}}$$

(12)

where the hermitian tensor $\underline{\underline{\mathbf{H}}}$ and the antihermiatian one $\underline{\underline{\mathbf{A}}}$ are defined by:

$$\underline{\underline{\mathbf{H}}} = \frac{\mathbf{Z}_s + \mathbf{Z}_s^\dagger}{2}, \qquad \underline{\underline{\mathbf{A}}} = \frac{\mathbf{Z}_s - \mathbf{Z}_s^\dagger}{2},$$

(13)

and then to use $\underline{\underline{\mathbf{A}}}$ as the impedance tensor. This choice leads to the formulation presented in [1]. However, this technique does not garentee a perfect match between (7) and (8) because a significant part of the tensor that ensures the exact phase (9) is neglected.

In this work, we introduce a second approach to overcome the above-mentioned problem. Equating the phase of the sum of the terms in (7) with the desired phase distribution (8) is quite complex because four impedance elements have to be controlled at the same time while maintaining the anti-hermitian property. In order to simplify this calculation, we take advantage of the fact that physical properties (energy conservation and reciprocity) are conserved under orthogonal transformations.

At each point of the metasurface we define a local framework such that the incident wave $\boldsymbol{\Psi}_{inc}$ is directed along a local axis (Fig. 2). Such an orthogonal reference system can be defined as:

$$\begin{aligned} \hat{\mathbf{x}}_2^{loc}(\boldsymbol{\rho}') &= \hat{\mathbf{h}}(\boldsymbol{\rho}') \\ \hat{\mathbf{x}}_1^{loc}(\boldsymbol{\rho}') &= \hat{\mathbf{x}}_2^{loc}(\boldsymbol{\rho}') \times \hat{\mathbf{n}} \end{aligned}.$$

(14)

Thus, denoting by $\underline{\mathbf{R}}(\boldsymbol{\rho}')$ the transformation matrix from the global to the local framework, the local wave and impedance can be written as

$$\begin{aligned} \mathbf{E}_t^{loc}(\boldsymbol{\rho}') &= \underline{\mathbf{R}}(\boldsymbol{\rho}') \cdot \mathbf{E}_t(\boldsymbol{\rho}') \\ \mathbf{H}_t^{loc}(\boldsymbol{\rho}') &= \underline{\mathbf{R}}(\boldsymbol{\rho}') \cdot \mathbf{H}_t(\boldsymbol{\rho}') \\ \underline{\underline{\mathbf{Z}}}_s^{loc}(\boldsymbol{\rho}') &= \underline{\mathbf{R}}(\boldsymbol{\rho}') \cdot \underline{\underline{\mathbf{Z}}}_s(\boldsymbol{\rho}') \cdot \underline{\mathbf{R}}(\boldsymbol{\rho}')^{-1} \end{aligned}$$

(15)

where the superscript $loc$ indicates that the quantity is written in the local framework.

In the local framework, eq.(3) becomes

$$\begin{bmatrix} E_1^{loc} \\ E_2^{loc} \end{bmatrix} = j \begin{bmatrix} X_{11}^{loc} & X_{12}^{loc} \\ X_{21}^{loc} & X_{22}^{loc} \end{bmatrix} \begin{bmatrix} -H_2^{loc} \\ 0 \end{bmatrix} = j \begin{bmatrix} X_{11}^{loc} & X_{12}^{loc} \\ X_{21}^{loc} & X_{22}^{loc} \end{bmatrix} \begin{bmatrix} J_1^{loc} \\ 0 \end{bmatrix}$$

(16)

As seen in equation (16), only the first column of $\underline{\underline{\mathbf{Z}}}_s^{loc}$ affects the electric field. The idea is to use eq. (11) in the local framework only for the elements of the first column, namely :

$$\begin{cases} X_{11}^{loc} = \bar{X}_{11}^{loc} \left[ 1 + M_{11}^{loc} \, \Im\left(\Psi_{obj,1}^{loc} \cdot \Psi_{inc,1}^{loc*}\right) \right] \\ X_{21}^{loc} = \bar{X}_{21}^{loc} \left[ 1 + M_{21}^{loc} \, \Im\left(\Psi_{obj,2}^{loc} \cdot \Psi_{inc,1}^{loc*}\right) \right] \end{cases}.$$

(17)

We can now ensure that $\underline{\underline{\mathbf{Z}}}_{loc}$ is anti-Hermitian by imposing $X_{12}^{loc} = X_{21}^{loc}$ as this component does not affect the electric field. In addition, $X_{22}^{loc}$ is a free quantity that we can use for other design purposes.

Finally, the impedance in the global framework that allows a phase match between the incident wave and the objective wave is obtained as

$$\underline{\underline{\mathbf{Z}}}_s \left( \boldsymbol{\rho}' \right) = \underline{\underline{\mathbf{R}}} \left( \boldsymbol{\rho}' \right) j \begin{bmatrix} X_{11}^{loc} \left( \boldsymbol{\rho}' \right) & X_{21}^{loc} \left( \boldsymbol{\rho}' \right) \\ X_{21}^{loc} \left( \boldsymbol{\rho}' \right) & X_{22}^{loc} \left( \boldsymbol{\rho}' \right) \end{bmatrix} \underline{\underline{\mathbf{R}}} \left( \boldsymbol{\rho}' \right)^{-1}. \quad (18)$$

As a final remark, it can be noted that eq.(18) has apparently 3 degrees of freedoms $\left( \bar{X}_{22}^{loc}, M_{11}^{loc}, M_{21}^{loc} \right)$. However, since $\bar{X}_{22}^{loc}$ affects the propagation constant $k_t^{sw}$, it cannot be arbitrarily selected.

### B. Amplitude control

Amplitude synthesis is based on a proper choice of the free parameters of eq.(18). As described in section II, the amplitude of the LW is proportional to the product $\bar{X}_{ij} M_{ij}$. Thus, the idea is to change this product along the metasurface in order to compensate the spreading factor of the SW and obtain an amplitude behavior proportional to that of the objective field.

As a result, the average reactance and the modulation index of each component of the impedance will vary depending on the position on the metasurface.

Eqs.(7),(8) imply that the following conditions have to be verified

$$\begin{cases} \bar{X}_{11}^{loc} \left( \boldsymbol{\rho}' \right) M_{11}^{loc} \left( \boldsymbol{\rho}' \right) \left| H_2^{loc} \left( \boldsymbol{\rho}' \right) \right| \propto \left| E_{obj,1} \left( \boldsymbol{\rho}' \right) \right| \\ \bar{X}_{12}^{loc} \left( \boldsymbol{\rho}' \right) M_{12}^{loc} \left( \boldsymbol{\rho}' \right) \left| H_2^{loc} \left( \boldsymbol{\rho}' \right) \right| \propto \left| E_{obj2} \left( \boldsymbol{\rho}' \right) \right| \end{cases} \quad (19)$$

Eq. (19) imposes the product $\bar{X}_{ij}^{loc} M_{ij}^{loc}$ but not their individual values. However, since the dispersion constant $k_t^{sw}$ is affected mainly by the average reactances values, smooth variations of these have to be imposed to avoid undesired reflections.

In the following, a linear variation of $\bar{X}_{ij}^{loc}$ along the direction of propagation is supposed. However, other variation laws could be used.

Equations (18),(19) represent the basic blocks for our metasurface antennas design procedure.

The most commonly used incident wave in antenna applications is the cylindrical SW generated by a coaxial probe placed at the center of the metasurface [1]-[2],[15]-[21]. In order to compensate the spreading factor of the incident wave, a radial variation of the tensor average impedance is imposed. This particular choice, introduces radial symmetry (invariance with respect to the azimuthal angle $\phi$) in $\bar{X}_{ij} \left( \boldsymbol{\rho}' \right) = \bar{X}_{ij} \left( \rho' \right)$, thus $\partial \bar{X}_{ij} \left( \rho' \right) / \partial \phi = 0$.

## IV. CYLINDRICAL WAVE EXCITATION CASE

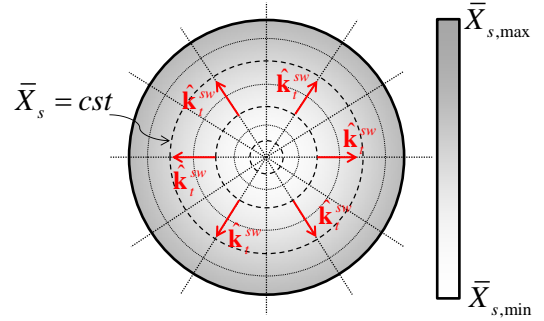

Fig. 3. Radial variation of the effective impedance over the metasurface.

As a result, the excited cylindrical surface wave travelling from the center to the perifery of the metasurface continues to have a radial direction of propagation $\hat{\mathbf{k}}_t^{sw} \left( \boldsymbol{\rho}' \right) = \hat{\boldsymbol{\rho}}, \forall \boldsymbol{\rho}'$ as shown in Fig.3.

For this important geometry, we use as a global framework the cylindrical coordinates system. Equation (3) becomes

$$\begin{bmatrix} E_\rho \left( \boldsymbol{\rho}' \right) \\ E_\phi \left( \boldsymbol{\rho}' \right) \end{bmatrix} = j \begin{bmatrix} X_{\rho\rho} \left( \boldsymbol{\rho}' \right) & X_{\rho\phi} \left( \boldsymbol{\rho}' \right) \\ X_{\rho\phi} \left( \boldsymbol{\rho}' \right) & X_{\phi\phi} \left( \boldsymbol{\rho}' \right) \end{bmatrix} \begin{bmatrix} -H_\phi \left( \boldsymbol{\rho}' \right) \\ H_\rho \left( \boldsymbol{\rho}' \right) \end{bmatrix} \quad (20)$$

The SW mode propagating on the metasurface will have the following form (APPENDIX (A1)-(A7)) :

$$\mathbf{E}_t \left( \boldsymbol{\rho}' \right) = I_{TM} \left( \boldsymbol{\rho}' \right) j \zeta_0 \left[ \Delta \left( \boldsymbol{\rho}' \right) \hat{\boldsymbol{\rho}} + \frac{X_{\rho\phi} \left( \boldsymbol{\rho}' \right) - \zeta_0 \Delta \left( \boldsymbol{\rho}' \right)}{\Delta \left( \boldsymbol{\rho}' \right) X_{\rho\phi} \left( \boldsymbol{\rho}' \right)} \hat{\boldsymbol{\phi}} \right] \mathrm{H}_1^{(2)} \left( k_{sw} \rho' \right)$$

$$\mathbf{H}_t \left( \boldsymbol{\rho}' \right) = -I_{TM} \left( \boldsymbol{\rho}' \right) \left[ \frac{X_{\rho\phi} \left( \boldsymbol{\rho}' \right) - \zeta_0 \Delta \left( \boldsymbol{\rho}' \right)}{X_{\rho\phi} \left( \boldsymbol{\rho}' \right)} \hat{\boldsymbol{\rho}} + \hat{\boldsymbol{\phi}} \right] \mathrm{H}_1^{(2)} \left( k_{sw} \rho' \right)$$

$$(21)$$

where $\zeta_0$ is the free space impedence, $\Delta$ is expressed in (A5) and the corresponding propagation constant $k_t^{sw}$ is given by (A6).

In order to have a surface wave $\Delta$ must be positive. However, depending on the values of $X_{\rho\rho}$, $X_{\rho\phi}$ and $X_{\phi\phi}$, four different cases are possible. 1) $X_{\rho\rho} X_{\phi\phi} > 0$: $\Delta$ only has a positive value, that means a SW type can propagate on the metasurface. 2) $\left( X_{\rho\rho} X_{\phi\phi} - \zeta_0^2 - X_{\rho\phi} X_{\phi\rho} \right)^2 > 4 \zeta_0^2 X_{\phi\phi} X_{\rho\rho} > 0$: two values are possible for $\Delta$. These values have the same sign as $X_{\phi\phi}$, giving rise to either no SW or two SW with different propagation constants $k_{t,1}^{sw}$ and $k_{t,2}^{sw}$. 3) $\left( X_{\rho\rho} X_{\phi\phi} - \zeta_0^2 - X_{\rho\phi} X_{\phi\rho} \right)^2 < -4 \zeta_0^2 X_{\phi\phi} X_{\rho\rho}$: $\Delta$ is complex, that does not correspond to any physical SW. 4) $\left( X_{\rho\rho} X_{\phi\phi} - \zeta_0^2 - X_{\rho\phi} X_{\phi\rho} \right)^2 = -4 \zeta_0^2 X_{\phi\phi} X_{\rho\rho}$, leading to $X_{\rho\phi} = 0$ and $X_{\rho\rho} X_{\phi\phi} = -\zeta_0^2$: $\Delta$ only has a positive value corresponding to a TE and a TM waves propagating independently with the same propagation constant $k_t^{sw}$.

In order to have only one excited surface wave on the metasurface, we opted to satisfy the first condition.

### A. Local framework derivation

The local framework associated to expression (21) is given by:

$$\hat{\mathbf{x}}_1^{loc}(\boldsymbol{\rho}') = \frac{1}{\sigma(\boldsymbol{\rho}')}\hat{\boldsymbol{\rho}} - \frac{\left(X_{\rho\rho}(\boldsymbol{\rho}') - \zeta_0\Delta(\boldsymbol{\rho}')\right)}{\sigma(\boldsymbol{\rho}')X_{\rho\phi}(\boldsymbol{\rho}')}\hat{\boldsymbol{\phi}}$$

$$\hat{\mathbf{x}}_2^{loc}(\boldsymbol{\rho}') = \frac{1}{\sigma(\boldsymbol{\rho}')}\hat{\boldsymbol{\phi}} + \frac{\left(X_{\rho\rho}(\boldsymbol{\rho}') - \zeta_0\Delta(\boldsymbol{\rho}')\right)}{\sigma(\boldsymbol{\rho}')X_{\rho\phi}(\boldsymbol{\rho}')}\hat{\boldsymbol{\rho}}$$

(22)

where $\sigma(\boldsymbol{\rho}') = \sqrt{1 + \left(X_{\rho\rho}(\boldsymbol{\rho}') - \zeta_0\Delta(\boldsymbol{\rho}')\right)^2 / X_{\rho\phi}^2(\boldsymbol{\rho}')}$.

The transformation matrix $\underline{\underline{\mathbf{R}}}(\boldsymbol{\rho}')$ is directly obtained from (22):

$$\underline{\underline{\mathbf{R}}}(\boldsymbol{\rho}') = \begin{bmatrix} \dfrac{1}{\sigma(\boldsymbol{\rho}')} & -\dfrac{X_{\rho\rho}(\boldsymbol{\rho}') - \zeta_0\Delta(\boldsymbol{\rho}')}{\sigma(\boldsymbol{\rho}')X_{\rho\phi}(\boldsymbol{\rho}')} \\ \dfrac{X_{\rho\rho}(\boldsymbol{\rho}') - \zeta_0\Delta(\boldsymbol{\rho}')}{\sigma(\boldsymbol{\rho}')X_{\rho\phi}(\boldsymbol{\rho}')} & \dfrac{1}{\sigma(\boldsymbol{\rho}')} \end{bmatrix}$$

(23)

The local incident current/magnetic field is then obtained using (23) in (15)

$$\mathbf{H}_i^{loc}(\boldsymbol{\rho}') = \underline{\underline{\mathbf{R}}}(\boldsymbol{\rho}')\cdot\mathbf{H}_i(\boldsymbol{\rho}') = -I_{TM}(\boldsymbol{\rho}')\sigma(\boldsymbol{\rho}')\mathrm{H}_1^{(2)}\left(k_t^{sw}\rho'\right)\hat{\mathbf{x}}_2^{loc}(\boldsymbol{\rho}')$$

$$\mathbf{J}^{loc}(\boldsymbol{\rho}') = \hat{\mathbf{n}}\times\mathbf{H}_i^{loc}(\boldsymbol{\rho}') = I_{TM}(\boldsymbol{\rho}')\sigma(\boldsymbol{\rho}')\mathrm{H}_1^{(2)}\left(k_t^{sw}\rho'\right)\hat{\mathbf{x}}_1^{loc}(\boldsymbol{\rho}')$$

(24)

and finally, the local *incident wave* needed in (17) is found as

$$\boldsymbol{\Psi}_{inc}^{loc}(\boldsymbol{\rho}') = \Psi_{inc,1}^{loc}(\boldsymbol{\rho}')\hat{\mathbf{x}}_1^{loc}(\boldsymbol{\rho}') = e^{jArg\left(I_{TM}(\boldsymbol{\rho}')\mathrm{H}_1^{(2)}\left(k_t^{sw}\rho'\right)\right)}$$

(25)

### B. Incident wave amplitude and phase estimation

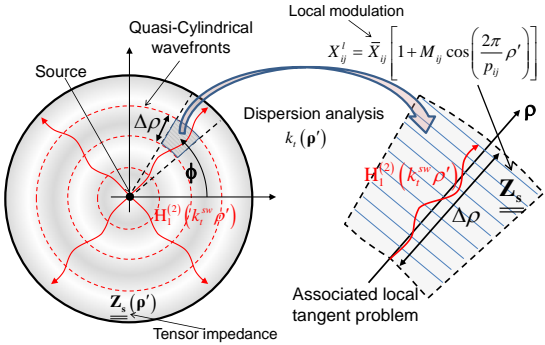

Fig. 4. Dividing the metasurface into sectorial areas with mono-directional propagation.

The magnetic field amplitude variation $I_{TM}(\boldsymbol{\rho}')$ needed in eqs.(21),(24),(25), can be numerically found using the Geometrical Optic (GO)-based procedure introduced in [24]. The idea is to consider the modulated impedance as a double-scale problem, with a fast variation (with respect to the wavelength) given by the sinusoidal modulation ($p_{ij} < \lambda$) and a slower variation given by the space dependence of the modulation parameters ($M_{ij}$, $\overline{X}_{ij}$, and $p_{ij}$). Under optic approximation, the fast impedance variation can be described by an equivalent impedance (homogenized problem) obtained using the local periodic dispersion. Thus, the modulated metasurface can be assimilated to a continuous slowly-variating equivalent surface impedance. It is important to note that even the metasurface is lossless; the equivalent impedance has a real part (resistance) describing the energy loss due to leaky wave radiation.

In order to numerically solve the propagation problem, the metasurface is divided into sectorial areas where propagation can be considered mono-directional. Then, each sectorial area is divided into various subdomains along the propagation direction, in which the impedance modulation (18) is assumed to be constant ($\overline{X}_{ij}$, $M_{ij}$ and $p_{ij}$ are constants).

The propagation problem within each subdomain can be analytically solved or approximated. The global solution is obtained by imposing the continuity of tangential field across each domain. As for optics or plane-waves this condition is equivalent to energy conservation across the interfaces.

In the case of cylindrical modulation, under small impedance modulation, the propagation can be still considered along the radial direction $\hat{\mathbf{k}}_i^{sw}(\boldsymbol{\rho}') = \hat{\boldsymbol{\rho}}, \forall \boldsymbol{\rho}'$.

For each direction $\phi$, an associated local tangent problem with radial direction of propagation can be defined as shown in Fig.4. Each local tangent problem is described by the surface impedance tensor $\underline{\underline{\mathbf{Z}}}_s(\boldsymbol{\rho}')$.

Each sectorial area is then divided in N sections where the $\boldsymbol{\rho}'$ depending quantitites can be considered constant. The incident magnetic field in the $n$-th section can be approximated in the local framework as

$$\mathbf{H}_i^{loc}(\boldsymbol{\rho}') = -I_n^{TM}(\boldsymbol{\rho}')\sigma(\boldsymbol{\rho}')\mathrm{H}_1^{(2)}\left(k_{t,n,\phi}^{sw}\rho'\right)\hat{\mathbf{x}}_2^{loc}(\boldsymbol{\rho}')$$

(26)

where $k_{t,n,\phi}^{sw} = \beta_{t,n,\phi}^{sw} - j\alpha_{t,n,\phi}^{sw}$ is obtained using the local tangent modulated impedance dispersion [25]. $\beta_{t,n,\phi}^{sw}$ is the real perturbed local wavenumber and $\alpha_{t,n,\phi}^{sw}$ is the local leakage attenuation parameter. The complex amplitude $I_{n,\phi}^{TM}$ is obtained by imposing the continuity of the field across two adjacent sectors as described by equation

$$I_{n,\phi}^{TM} = \frac{I_{n-1,\phi}^{TM}\sigma^{n-1}(\rho_n')\mathrm{H}_1^{(2)}\left(k_{t,n-1,\phi}^{sw}\rho_n'\right)}{\sigma^n(\rho_n')\mathrm{H}_1^{(2)}\left(k_{t,n,\phi}^{sw}\rho_n'\right)}$$

(27)

where $\rho_n'$ is the position at the interface.

Imposing the continuity of the tangential field (amplitude and phase) along the propagation path starting from the center of the metasurface to the periphery leads to the numerical integration of the phase of the incident wave.

As a final remark; we want to stress the fact that it is important to minimize the gradient of the GO equivalent refraction index $\eta_{eq} = k_t^{sw}/k_0$ [24] in order to keep, as much as possible, wave propagation along the radial axis. This latter quantity is given by:

$$\eta_{eq}(\boldsymbol{\rho}') = \sqrt{1 - \left[\frac{jX_{\rho\rho}(\boldsymbol{\rho}')}{\zeta} + \frac{X_{\rho\phi}^2(\boldsymbol{\rho}')k_z^{sw}(\boldsymbol{\rho}')}{\zeta^2 k_0 + j\zeta X_{\phi\phi}(\boldsymbol{\rho}')k_z^{sw}(\boldsymbol{\rho}')}\right]^2}$$

(28)

where $k_z^{sw}(\boldsymbol{\rho}') = \sqrt{k_0^2 - \left(k_t^{sw}(\boldsymbol{\rho}')\right)^2}$ .

Thus, the free quantity $X_{22}^{loc}$ in (18) has to be selected in order to minimize the gradient $\nabla \eta_{eq}$ of (28).

## V. METASURFACE AND APERTURE FIELD ANTENNA DESIGN

### A. Impedance design

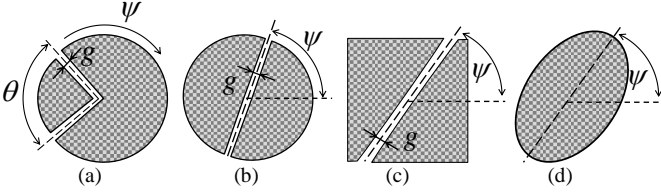

Fig. 5. Unit cell topologies. The chosen cell (a) consists of a circular patch with a v shaped slot. It is defined by slot wideness $g$ , the orientation angle $\psi$ and the slot aperture angle $\theta$ .

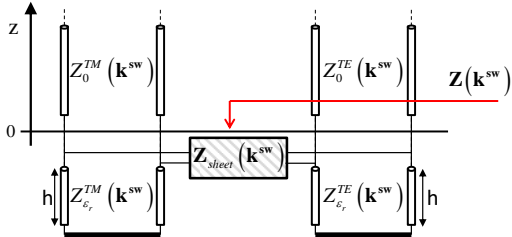

Fig. 6. Printed metasurface equivalent TM and TE transmission lines.

The obtained surface impedance law $\underline{\mathbf{Z}_s}(\boldsymbol{\rho}')$ can be implemented by printing asymmetric subwavelength patches in a squared lattice over a dielectric substrate as already proposed in [1] and [20]-[22].

An in-house tool based on periodical Method of Moment is used to characterize the surface impedance taking into account all the possible dispersion effects (dielectric slab dispersion, etc..). First, the resonating wavenumber $\mathbf{k}^{sw}$ associated to the dominant guided wave is found. Then, the sheet impedance $\mathbf{Z}_{sheet}(\mathbf{k}^{sw})$ (see Fig.6) describing the patch array at this particular wavenumber is calculated. Finally, the impenetrable impedance $\mathbf{Z}(\mathbf{k}^{sw})$ is obtained by adding the contribution of the dielectric slab at the resonant wavenumber.

Several patch topologies were investigated in order to find the geometry able to ensure the highest reactance variation range. In order to have independent control of reactance component variations, a large range of impedance values is needed. This can be obtained by adding degrees of freedom to the cell geometry. As a result, a new geometry, composed of a circular patch with a variable v-shaped slot (Fig. 5a) is introduced. This cell presents only one axe of symmetry, however in contrast with what is stated in [26] its equivalent tensor is a purely imaginary symmetric tensor. This property is a direct consequence of the absence of losses and of the reciprocity of the used media [23].

This geometry can produce variable cross-reactance values acting mainly on the slot opening angle $\theta$ . This additional degree of freedom leads to a four-dimensional database

(incident angle, patch dimension, patch orientation angle $\psi$ , and slot opening angle $\theta$ ) that will significantly increase the design possibilities offered by this unit cell.

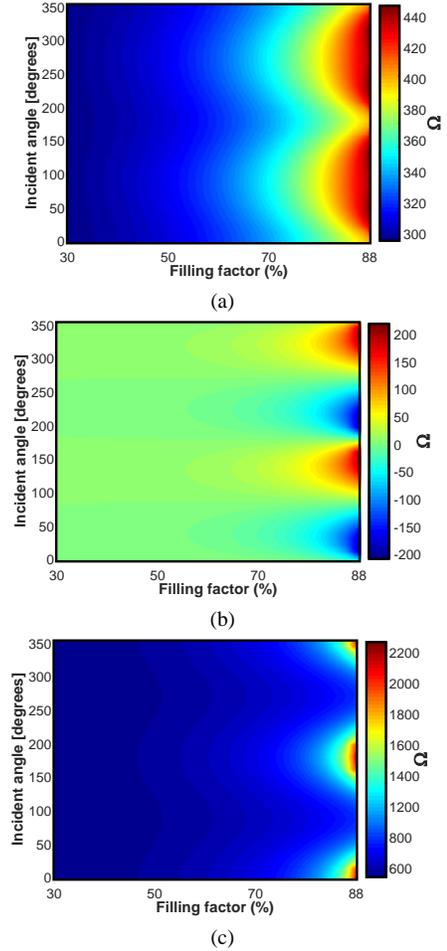

Fig. 7. Reactance levels for the components $X_{\rho\rho}$ (a), $X_{\phi\rho}$ (b), and $X_{\phi\phi}$ (c).

It is important to stress that, since the cell is asymmetrical and the lattice is squared, a rotation $\psi$ of the cell is not equivalent to a change of the incident angle of the SW on the patch lattice. As a result, impedance depends on the orientation angle $\psi$ .

An example of impedance values obtained with the new patch geometry at $20\,\mathrm{GHz}$ is shown in Fig.7. The periodic cells of dimension $\lambda_0/12$ , where $\lambda_0$ is the free space wavelength, are printed on a Rogers TMM6 substrate with relative permittivity $\varepsilon_r = 6$ and thickness $h = 1.27\,\mathrm{mm}$ . The plots show the impedance as a function of the incident angle (vertical axis) and the ratio between the patch radius and the lattice dimensions (horizontal axis) for $\theta = 60°$ and $\psi = 60°$ .

A database containing tensorial reactance maps (similar to those in Fig.7) was generated for different geometrical configurations (orientation angle $\psi$ , slot aperture angle $\theta$ , slot wideness) and used to implement the desired reactance levels in the following antenna examples. The obtained reactance values $X_{\rho\rho}, X_{\phi\rho}, X_{\phi\phi}$ belong to the following ranges:

- $280\Omega < X_{\rho\rho} < 450\Omega$
- $-350\Omega < X_{\phi\rho} < 350\Omega$
- $400\Omega < X_{\phi\phi} < 2900\Omega$

### B. Aperture field metasurface antenna design

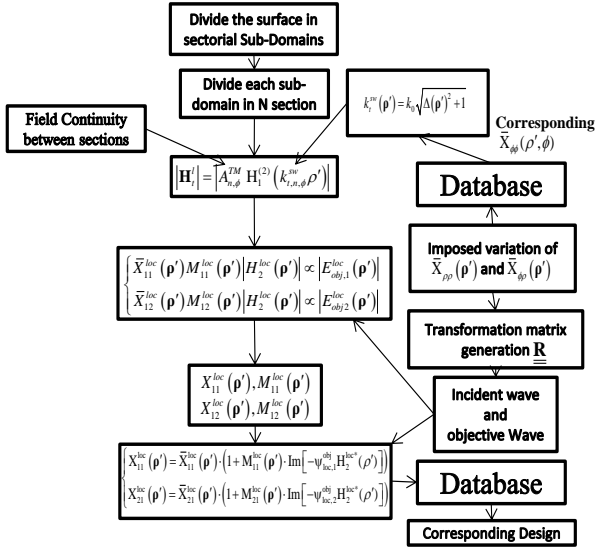

Fig. 8. Metasurface implementation algorithm.

Fig.8 summarizes the design steps of a tensorial metasurface able to implement a given objective electric field distribution $\mathbf{E}_t^{obj}(\boldsymbol{\rho}')$. Two preliminary steps are needed:

- Step 1: The average impedances variations laws and the maximum values of the modulation indexes are found by ensuring that the reactance values are available in the database.

- Step 2: The incident surface wave is evaluated by decomposing the metasurface into various subdomains along the propagation direction, in which the surface modulation is assumed to be constant. For a cylindrical SW the metasurface is devided into angular sub-domains, each sub-domaine is then devided into N elementary radial sections.

Next, starting from the source and following the propagation, the following steps are performed on each sector:

- Step 3: The propagation constant $k_{t,n,\theta}^{sw}$ is found by solving the local dispersion problem, then amplitude and phase of the magnetic field are found via (26) imposing tangential field continuity at sector interface.

- Step 4: The transformation matrix $\underline{\mathbf{R}}(\boldsymbol{\rho}')$ is found, the incident and the objective fields are then written in the local framework.

- Step 7: Modulation indexes $M_{11}^{loc}(\boldsymbol{\rho}')$ and $M_{12}^{loc}(\boldsymbol{\rho}')$ are calculated using equation (19).

- Step 8: Impedance variations are obtained using equations (18) and (19).

- Step 9: The metasurface is discretized according to the selected periodic lattice. Following this, the metasurface layout is obtained by selecting the corresponding patch geometry available in the database.

## VI. NUMERICAL EXAMPLES

This section presents several examples of aperture field implementation obtained via the proposed method. Since, the goal of this section is to validate our approach; four analytical expressions of the aperture field have been selected in order to radiate respectively: a LP beam; a CP beam; a multi-beam diagram having 4 beams with different polarizations; and a CP flat-top diagram.

For real applications, numerical optimization procedures could be used in order to find the optimal aperture distribution.

The metasurfaces were analyzed using commercial full-wave software. However, the complexity of the geometrical details severely limits the size of the antenna that can be analyzed.

The numerical results were validated by comparison respectively with:

1) the theoretical far field of the aperture distribution (proportional to the product between the Fourier transform of the aperture field component and the corresponding magnetic dipole pattern);

2) the free space radiation of the equivalent magnetic current of the theoretical LW field (7), namely $\mathbf{J}_{LW}(\boldsymbol{\rho}') = 2\hat{\mathbf{n}} \times \mathbf{E}_t^{LW}$.

### A. Single beam linearly polarized antenna:

A single beam metasurface antenna that radiate a linearly polarized single beam pointing at $(\theta_0, \phi_0)$ can be obtained by imposing the following objective aperture field:

$$\mathbf{E}_t^{obj}(\boldsymbol{\rho}') = e^{-jk_0(\sin\theta_0\cos\phi_0 x' + \sin\theta_0\sin\phi_0 y')}\hat{\mathbf{e}}(\phi_0) \quad (29)$$

where the amplitude is constant over the aperture and the polarization of beam is controlled by $\hat{\mathbf{e}}(\phi_0)$ as :

$$\hat{\mathbf{e}}(\phi_0) = \begin{cases} \cos\phi_0\hat{\mathbf{x}} + \sin\phi_0\hat{\mathbf{y}} & \text{for }\hat{\boldsymbol{\theta}}\text{ polarization} \\ -\sin\phi_0\hat{\mathbf{x}} + \cos\phi_0\hat{\mathbf{y}} & \text{for }\hat{\boldsymbol{\phi}}\text{ polarization} \end{cases} \quad (30)$$

We first considered the case of a circular aperture pointing at $(\theta_0, \phi_0) = (30°, 0°)$ with polarization along the phi axis ( $\hat{\mathbf{e}} = \hat{\mathbf{y}}$ ). The radius of the aperture was limited to $3\lambda$ in order to perform the electromagnetic analysis using two commercial softwares.

The design was obtained using the proposed algorithm imposing constant values of the average impedance $(\overline{X}_{\rho\rho}, \overline{X}_{\phi\rho}, \overline{X}_{\phi\phi}) = (360, 80, 1500)\Omega$, while the corresponding maximum values of the modulation indexes in the local framework were equal to $(M_{11,max}^{loc}, M_{12,max}^{loc}) = (0.2, 0.45)$. The final antenna layout is composed of 3720 patches and it is shown in the inset of Fig.9a.

The antenna was analyzed using two commercial software: Ansys Designer$^{TM}$ and FEKO$^{TM}$. However, in order to perform the simulation with FEKO on a workstation with 128GB of RAM the model was simplified by using a poorer-quality mesh with respect to Designer.

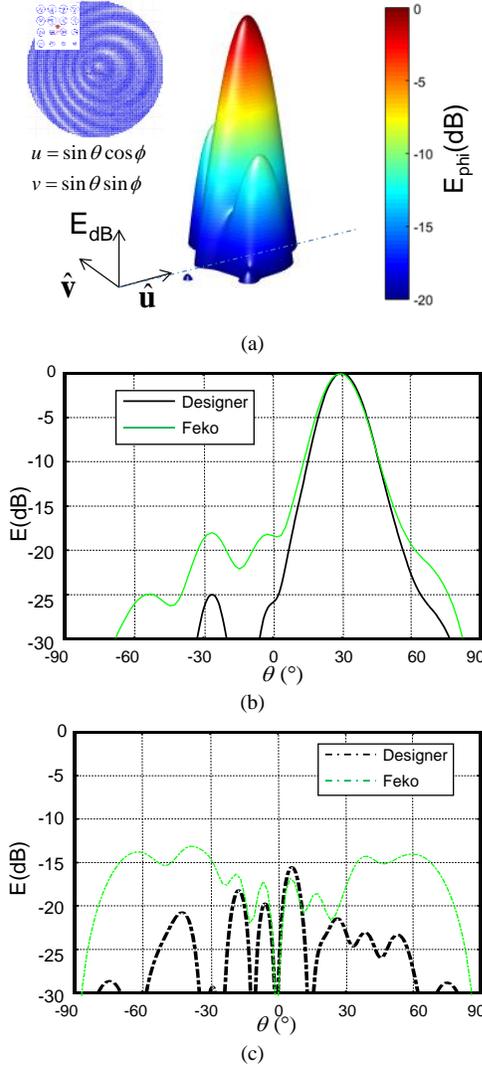

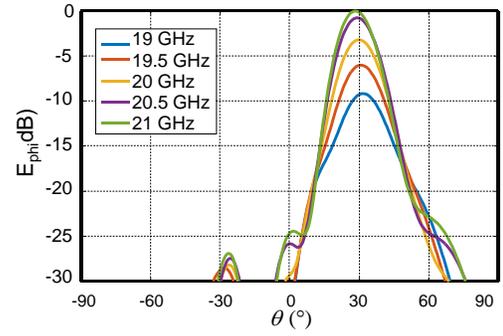

Fig. 10. Phi component of the far field radiation pattern at the $\phi = 0°$ cut plane for different frequencies

Fig.10 represents the phi component of the radiated field in the frequency range $[19, 21]$ GHz. As for any LW antennas, the beam direction deviates from the design direction $\theta_0$ as the frequency changes. This is due to dispersion effects on the propagation constants $k_{sw}$ and on tensor impedance elements. Heretofore and in order to consider larger apertures, only results obtained with Ansys Designer will be presented.

### B. Broadside circularly polarized antenna:

A CP beam can be obtained by superposing two distributions of the form (29) radiating orthogonal LP beams with a $\pi/2$ phase shift. Thus, a broadside Right Hand CP (RHCP) antenna can be generated by distribution (29) with a normalized polarization vector given by

$$\hat{\mathbf{e}}(\theta_0) = \left(1/\sqrt{2}\right)\left(\hat{\mathbf{x}} + j\hat{\mathbf{y}}\right). \tag{31}$$

A circular aperture with radius $5\lambda$ was considered. Using the same impedance parameters as for the previous example, an antenna layout composed of 10732 patches was obtained (see inset of Fig.11(a)).

Fig. 11 presents the far field results. As expected, the radiation pattern presents a broadside CP beam as shown in the 3D pattern (Fig 11(a)). Figures 11(b) and 11(c) show the comparison in the $\phi = 0°$ cut-plane between the field obtained with the theoretical aperture radiation (blue line) and the field radiated by the equivalent current (red lines) for the CRH and CLH, respectively.

The curves are similar but a higher SLL is obtained with the metasurface implementation. This difference is due to the fact that some of the energy associated with the surface wave reaches the edge of the antenna and is reflected backwards (radiating in the wrong direction). Using a larger metasurface, it is easier to make sure that maximum energy associated with the SW is radiated in order to prevent undesired edge reflections.

Fig. 9. Far field radiation (normalized) results for the metasurface implementation of distribution (29) pointing at $(\theta_0, \phi_0) = (30°, 0°)$: (a) 3D representation of the radiation pattern; (b) Phi component in the $\phi = 0$ cut-plane. (c) Theta component in the $\phi = 0$ cut-plane.

Fig.9 presents the far field results obtained using Ansys Designer$^{TM}$ and FEKO$^{TM}$. A 3D representation of the radiation pattern is described in Fig.9(a), while the $\phi = 0°$ cut plane is presented in Fig.9(b) and (c). The solid lines correspond to the phi component while the dotted lines correspond to the theta component.

As expected, the radiation pattern presents a tilted beam pointing at the desired direction and having the predicted polarization. The cross-polarization level differs between the two commercial software (-22dB Designer and -18.5dB FEKO). This discrepancy is probably due to the varying-quality mesh that was used in the two simulations.

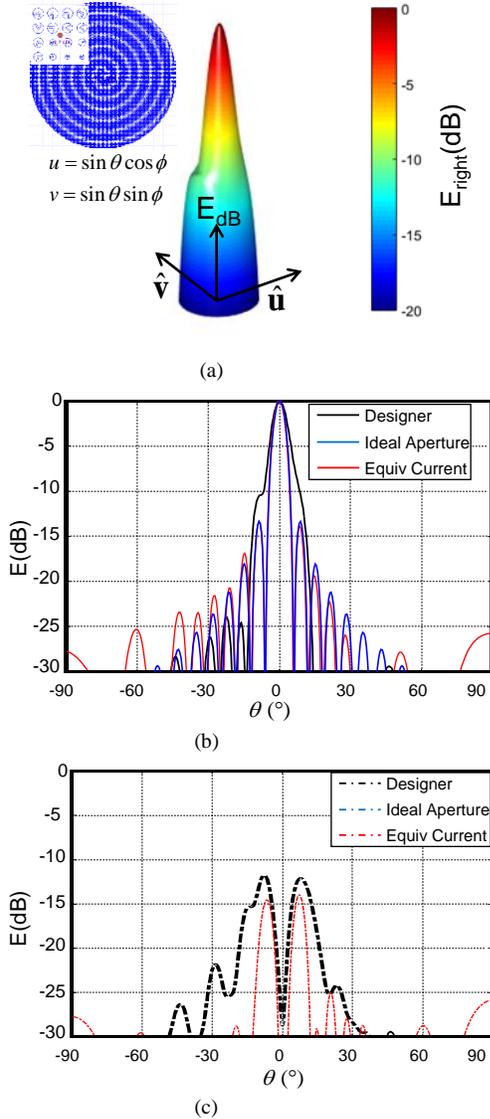

(a)

(b)

(c)

Fig. 11. Far field radiation pattern (normalized) (a) 3D representation the radiation pattern (designer). (b) RHCP component in the $\phi = 0°$ cut plane. (c) LHCP component in the $\phi = 0°$ cut plane.

## C. Multi-beam antenna:

As a third example we consider a multi-beam antenna (4 beams). A scalar metasurface able to radiate four beams is already presented in the literature [27]. In this work, the desired radiation pattern is obtained by dividing the metasurface into four regions (one for each beam). In our example, the whole surface is used in order to generate the four radiation beams. In addition, the polarization of each beam can be independently controlled by acting on the modulation of the reactance component.

To this goal, an aperture field distribution that radiates multiple beams can be obtained by superposing distributions of the form (29), (30). We selected a 4-beam pattern with the following characteristics:

- beam 1: LP along $\theta$, $(\theta_0, \phi_0) = (30°, 0°)$;

- beam 2: LP along $\phi$, $(\theta_2, \phi_2) = (45°, 180°)$;

- beam 3: RHCP, $(\theta_3, \phi_3) = (45°, 270°)$;

- beam 4: LHCP, $(\theta_4, \phi_4) = (30°, 90°)$;

This choice is motivated by the fact that the beams have all the possible polarization states.

The corresponding aperture field distribution in cylindrical components is depicted in Fig.12. It is noteworthy that strict control of the amplitude and phase distribution is needed to obtain the desired multi-beam radiation.

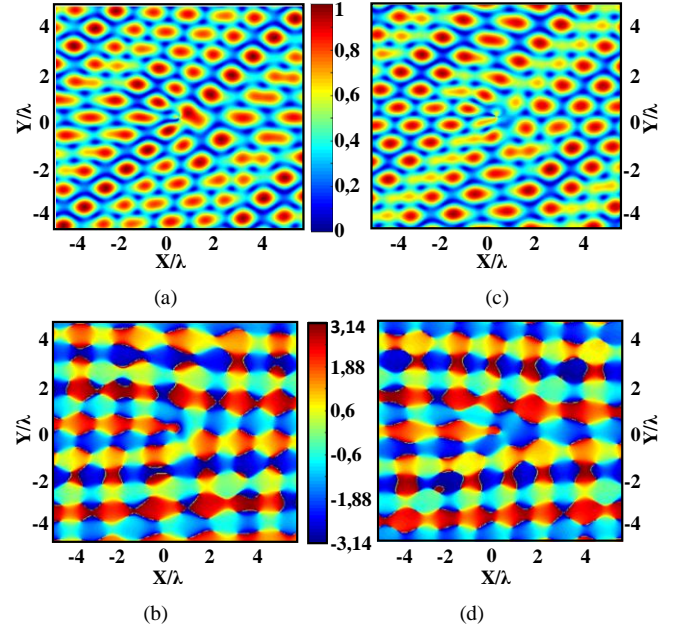

(a)                                                    (c)

(b)                                                    (d)

Fig. 12. Aperture field distribution of the 4 beams metasurface antenna.(a) $\left|E_\rho^{obj}\right|$; (b) $\arg(E_\rho^{obj})$ (c) $\left|E_\phi^{obj}\right|$; (d) $\arg(E_\phi^{obj})$.

Fig.13 presents the far field radiation pattern. A 3D representation of the total electric field obtained with Designer is illustrated in Fig.13(a). Fig.13(b) and (c) present the circular components of the far field in the $\phi = 90°$ cut-plane while Fig.13(d) and (e) present the linear components of the far field in the $\phi = 0°$ cut-plane. As expected, the radiation pattern presents four beams pointing in the desired directions with the expected polarization. Moreover, a close agreement between the numerical and theoretical fields (theoretical aperture radiation *blue line* and the field radiated by the equivalent currents *red lines*) is achieved.

Fig.14 shows the far field radiation pattern at frequencies 19.5, 20, and 20.5 GHz. It can be seen that the antenna performances (polarization and side lobe levels) are acceptable over the considered frequencies.

These results, in light of the selected complex aperture distribution, confirm that the proposed method is able to implement general radiating aperture distribution.

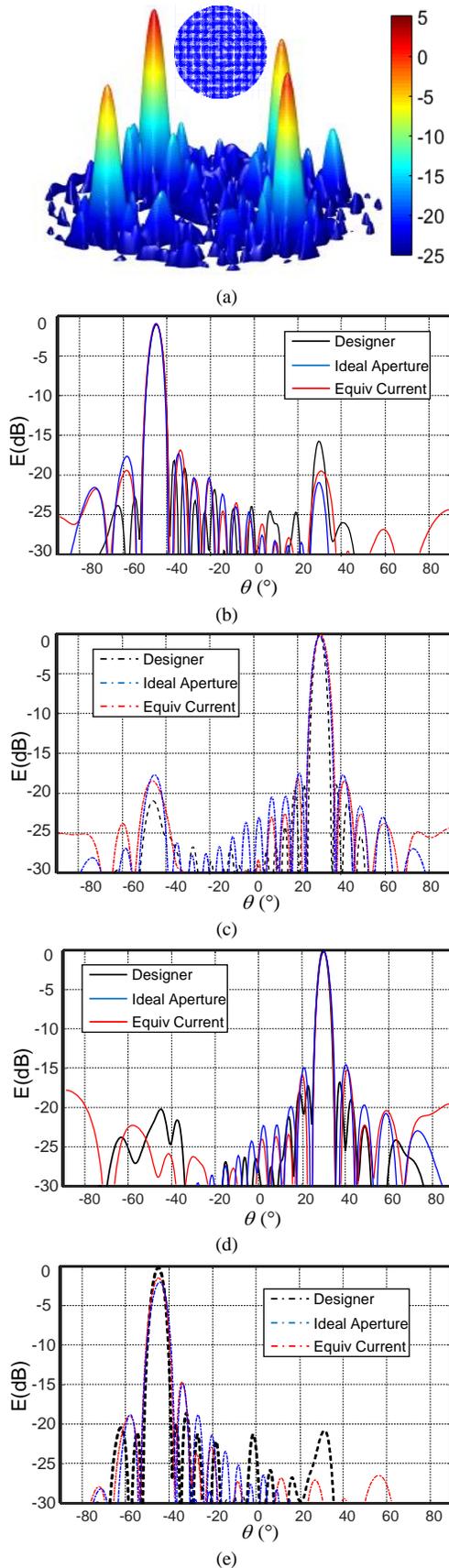

Fig. 13. Multi-beam metasurface radiation pattern (normalized) (a) 3D radiation pattern (Designer). (b) RHCP component in the $\phi = 90°$ cut-plane. (c) LHCP component in the $\phi = 90°$ cut-plane. (d) Theta component in the $\phi = 0°$ cut-plane. (e) Phi component in the $\phi = 0°$ cut-plane.

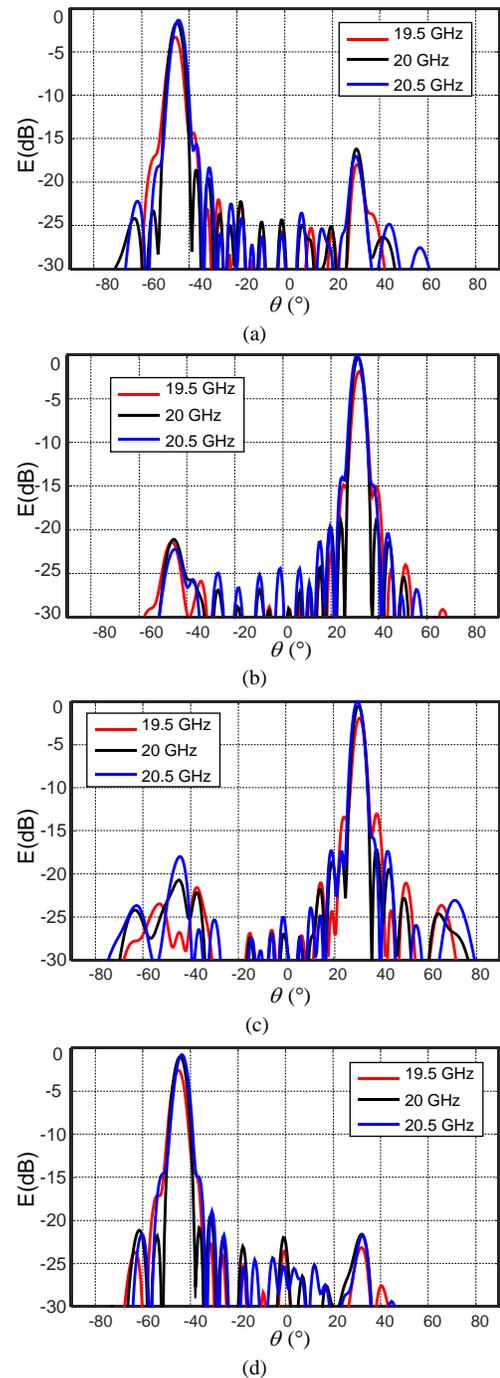

Fig. 14. Multi-beam metasurface radiation pattern (normalized) at 19.5 GHz (red line) 20 GHz (black line) and 20.5 GHz (blue line). (a) RHCP component in the $\phi = 90°$ cut-plane. (b) LHCP component in the $\phi = 90°$ cut-plane. (c) Theta component in the $\phi = 0°$ cut-plane. (d) Phi component in the $\phi = 0°$ cut-plane.

### D. Flat top circularly polarized antenna:

As a final example, we consider a flat-top RHCP antenna. For this kind of antenna, the normalized radiation pattern $F_n$ is almost constant within an angular region, while it is below a certain threshold outside of the region. Namely

$$\|F_n(\theta)\| = \begin{cases} 1 \pm \delta & \text{if } |\theta| < \theta_f \\ < \tau & \text{if } |\theta| > \theta_f \end{cases} \qquad (32)$$

where $\delta$ is the maximum oscillation with respect to the desired level and $\tau < 1$ is the threshold.

An aperture distribution that can produce a radiated field with amplitude of the form (32) can be obtained via far-field to near-field transformation using the bi-dimensional inverse Fourier transformation of the characteristic function of a circular disk. The final aperture field radiating a RCHP wave is finally obtained as

$$\mathbf{E}_t^{obj}(\boldsymbol{\rho'}) = \frac{1}{\sqrt{2}}(\hat{\mathbf{x}} + i\hat{\mathbf{y}})\frac{J_1(k_0\rho'\sin\theta_f)}{k_0\rho'\sin\theta_f} \qquad (33)$$

As seen in (33), $\theta_f$ is fixed a priori, while quantities $\delta$ and $\tau$ in (32) depend on aperture dimension.

A radiation pattern with $\theta_f = 20°$ was selected. The corresponding theoretical oscillation and threshold are $\delta = \pm 1.1$dB and $\tau = -20$dB

The design algorithm was used with the following impedance parameters :

- $(\bar{X}_{\rho\rho}, \bar{X}_{\phi\phi}) = (360, 1500)\,\Omega$ .

- Linear variation of $\bar{X}_{\phi\rho}$ between $100\,\Omega$ and $190\,\Omega$ .

- $(M_{11,\max}^{loc}, M_{12,\max}^{loc}) = (0.2, 0.45)$ .

Fig. 15 presents the obtained radiation pattern with the antenna layout shown in the inset. A 3D representation of the radiation diagram simulated with Designer is depicted in Fig.15(a) while the $\phi = 0°$ cut-plane is reported for the RHCP and LHCP polarization in Fig 15(b) and (c) respectively.

As can be seen, the antenna has a flat top radiation pattern for $\theta < \theta_f$ with a RHCP and oscillation $\delta = \pm 1$dB in close agreement with the theoretical results.

## VII. CONCLUSION

A new procedure to design tensorial metasurface antennas able to implement general radiating aperture field distribution was presented. The phase synthesis is based on local holography while amplitude is obtained using variable impedance modulation (average impedances and modulation indexes).

The theory was initially validated with the design of simple aperture field distribution as a LP and CP beam antenna and afterwards, with more sophisticated designs (multibeam and flat-top antennas). A good agreement was obtained between the expected results and the numerical simulations.

## APPENDIX

This appendix presents the derivation of the hybrid EH SW guided by tensorial impedances boundary conditions of the form (1) with constant elements.

Proceeding as in [28], the electromagnetic field on the surafce is written as:

$$\begin{cases} \mathbf{E}_t = \left[V_{TM}\hat{\boldsymbol{\rho}} - V_{TE}\hat{\boldsymbol{\phi}}\right]\mathrm{H}_1^{(2)}\left(k_t^{sw}\rho'\right) \\ \mathbf{H}_t = -\left[I_{TM}\hat{\boldsymbol{\phi}} + I_{TE}\hat{\boldsymbol{\rho}}\right]\mathrm{H}_1^{(2)}\left(k_t^{sw}\rho'\right) \end{cases} \qquad (A1)$$

where $V_{TM}$, $V_{TE}$, $I_{TM}$ and $I_{TE}$ are the modal coefficients of the surface wave and $k_t^{sw}$ is the corresponding wave number. $\mathrm{H}_1^{(2)}$ is the first order Hankel Function of the second kind.

The impedance boundary condition introduces a relation between the modal coefficients:

$$\begin{bmatrix} V_{TM} \\ V_{TE} \end{bmatrix} = \begin{bmatrix} jX_{\rho\rho} & -jX_{\rho\phi} \\ -jX_{\rho\phi} & jX_{\phi\phi} \end{bmatrix}\begin{bmatrix} I_{TM} \\ I_{TE} \end{bmatrix} \qquad (A2)$$

We introduce the quantity $\Delta$ defined as

$$\Delta = \sqrt{\left(\frac{k_t^{sw}}{k_0}\right)^2 - 1} \qquad (A3)$$

Since for surface waves $k_t^{sw} > k_0$ , it follows that $\Delta > 0$ .

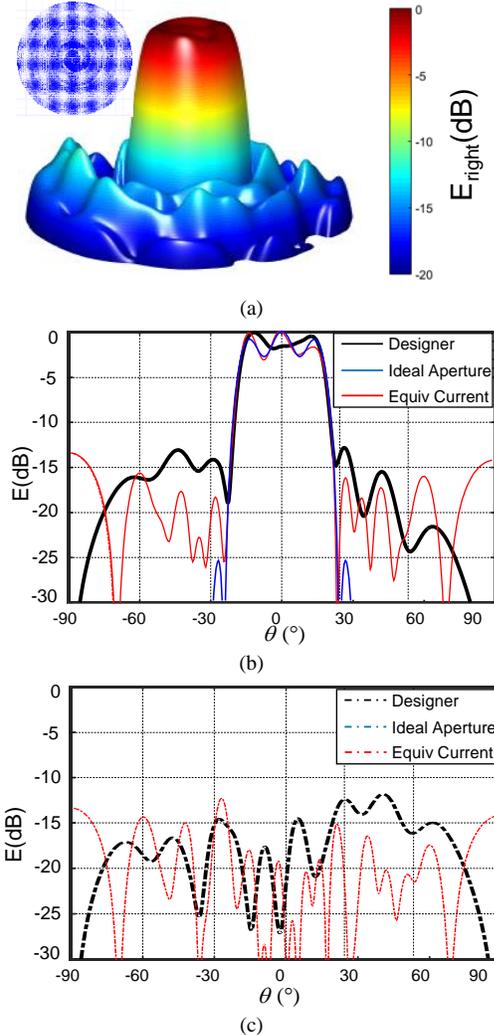

Fig.15. Far field radiation pattern (normalized) (a) 3D RHCP component of the electric field (Designer). (b) $\mathrm{E}_{right}$ comparison for the $\phi = 0$ cut plane . (c) $\mathrm{E}_{left}$ comparison for the $\phi = 0$ cut plane. The solid line and the dashed line represent the RHCP and the LHCP components respectively

The propagation constant $k_t^{sw}$ can be found using transverse resonance technique along the z axis leading to the following equation:

$$\Delta^2\left(\zeta_0 X_{\phi\phi}\right)+\Delta\left(-X_{\rho\rho}X_{\phi\phi}+\zeta_0^{\,2}+X_{\rho\phi}^{\,2}\right)-X_{\rho\rho}\zeta_0=0 \quad (A4)$$

The expression of $\Delta$ is obtained as:

$$\Delta\left(\boldsymbol{\rho'}\right)=\frac{\left(X_{\rho\rho}\left(\boldsymbol{\rho'}\right)X_{\phi\phi}\left(\boldsymbol{\rho'}\right)-\zeta_0^{\,2}-X_{\rho\phi}^{\,2}\left(\boldsymbol{\rho'}\right)\right)}{2\zeta_0 X_{\phi\phi}\left(\boldsymbol{\rho'}\right)}$$
$$\pm\frac{\sqrt{\left(X_{\rho\rho}\left(\boldsymbol{\rho'}\right)X_{\phi\phi}\left(\boldsymbol{\rho'}\right)-\zeta_0^{\,2}-X_{\rho\phi}^{\,2}\left(\boldsymbol{\rho'}\right)\right)^2+4\zeta_0^{\,2}X_{\phi\phi}\left(\boldsymbol{\rho'}\right)X_{\rho\rho}\left(\boldsymbol{\rho'}\right)}}{2\zeta_0 X_{\phi\phi}\left(\boldsymbol{\rho'}\right)}$$
$$(A5)$$

It is important to note that the physical solutions of (A5) are those that satisfy $\Delta>0$.

Finally, the corresponding propagation constant $k_t^{sw}$ is obtained as

$$k_t^{sw}\left(\boldsymbol{\rho'}\right)=k_0\sqrt{\Delta\left(\boldsymbol{\rho'}\right)^2+1} \quad (A6)$$

while the modal coefficients are given by:

$$\begin{cases} I_{TE}=\dfrac{\left(X_{\rho\rho}-\zeta_0\Delta\right)}{X_{\rho\phi}}I_{TM} \\[2ex] I_{TE}=\dfrac{X_{\phi\phi}}{\left(X_{\phi\phi}+\dfrac{\zeta_0}{\Delta}\right)}I_{TM} \end{cases}. \quad (A7)$$

27-Jun-2016

Dear Dr. Massimiliano Casaletti,

We have received a manuscript via Manuscript Central entitled "Implementation of radiating aperture field distribution using tensorial metasurfaces" for which you are listed as a co-author.  This manuscript is presently being given full consideration for publication.

Your manuscript ID is AP1606-0896.

Please refer to the above manuscript ID in all future correspondence.

If there are any changes to your affiliation or e-mail address, please visit Manuscript Central and edit your information.  This will ensure that we can contact you in a timely manner regarding the status of your submission.

According to the IEEE Publication Services and Products Board (PSPB) Operations Manual, it is the responsibility of authors to disclose if the manuscript has been published previously, is still under active consideration by another publication, or is closely related to any submitted or published conference or journal paper. Authors should include any closely related work as "supporting document" in the submission record if it is not available in the public domain. Please contact the Editorial Office immediately if you have not disclosed the above information in your submission record.

Thank you for submitting your manuscript to the IEEE Transactions on Antennas and Propagation.  By working together, we can ensure that the Transactions remains a high-quality publication.

Yours sincerely,

Miss Sunny Tse
Editorial Assistant
AP Transactions Website: http://ieeeaps.org/aps_trans/index.htm
AP Manuscript Central Website: https://mc.manuscriptcentral.com/tap-ieee